\documentclass[prd,amsfonts,nofootinbib,preprintnumbers,
showpacs,showkeys]{revtex4}
\usepackage{graphics}
\usepackage{epsfig}
\usepackage{latexsym}
\usepackage{amssymb, amscd}
\newcommand{\ds}{\displaystyle}

\newcommand{\mat}{\left ( \begin{array}}
\newcommand{\emat}{\end{array} \right )}
\newcommand{\vect}{\left ( \begin{array}{c}}
\newcommand{\evect}{\end{array} \right )}

\preprint{HU-EP-08/06}

\begin{document}

\title{ \bf Pion condensation of quark matter in the static Einstein
universe}
\author{D.~Ebert}
\affiliation{Institute of Physics, Humboldt-University , 12489
Berlin, Germany}
\author{K.G.~Klimenko}
\affiliation{Institute
of High Energy Physics, 142281, Protvino, Moscow Region, Russia}
\author{A.V.~Tyukov}
\author{V.Ch. Zhukovsky}
\affiliation{Faculty of Physics, Moscow State University, 119991,
 Moscow, Russia}

\begin{abstract}
In the framework of an extended Nambu--Jona-Lasinio model
we are studying pion condensation in quark matter with an asymmetric
isospin composition in a gravitational field of the static Einstein
universe at finite temperature and chemical potential. This
particular choice of the gravitational field configuration enables us
to investigate  phase transitions of the system with exact
consideration of the role of this field in the formation of quark and
pion condensates and to point out its influence on the phase
portraits. We demonstrate the effect of oscillations of the
thermodynamic quantities as functions of the curvature and also refer
to a certain similarity between the behavior of these quantities as
functions of curvature and finite temperature. Finally, the role of
quantum fluctuations for spontaneous symmetry breaking in the case of
a finite volume of the universe is shortly discussed.
\end{abstract}


\maketitle
\section{Introduction}

Low-energy nonperturbative effects in quantum chromodynamics (QCD),
especially  at nonzero densities,
can only be studied by approximate methods within the framework of
various effective models. It is well known that light meson physics
is described by four-fermion models, such as the Nambu--Jona-Lasinio
(NJL) model, which was successfully used to deal with dynamical
chiral symmetry breaking (D$\chi$SB) both in the vacuum and in
hot/dense baryonic matter (see, e.g. \cite{[1],[2]}, as well as the
reviews \cite{volkov,hatsuda} and references therein).
Recently, much attention has been paid to the effects of
diquark condensation and color superconductivity (CSC). The first
studies of the gap equations and the Ginzburg-Landau free energy for
a system of relativistic fermions led to the conclusion that
superconductive and color superconductive states may arise in
baryonic matter \cite{[9],[10]} (see also the recent papers
\cite{alford,bekvy}). Another interesting phenomenon, the
condensation of charged pions, which may  appear in dense hadronic
matter due to an asymmetry of its isospin composition, has been
investigated in the framework of QCD-like effective models, including
the NJL model, as well \cite{son,[15],[23],23,[18],[17],[19]}.

Note that all these phenomena might be inherent to physics of compact
stars, where rather strong magnetic as well as gravitational fields
are present. Therefore, investigations of the influence of an
external (chromo-)magnetic field on the properties of the  D$\chi$SB
phase transition \cite{oscil,[3]}, color superconductivity \cite{3}
and pion condensation \cite{[21]} effects are quite  
appropriate. In particular, it was shown in \cite{oscil,[3],3,[21]}
that external fields significantly change the properties of the
chiral and CSC phase transitions. In several papers, in the framework
of the NJL model, the influence of a gravitational field on the 
D$\chi$SB due to the creation of a finite quark condensate
$\langle\bar{q}q\rangle$ has been investigated at zero values of
temperature and chemical
potential~\cite{MUTA,Elizalde,Elizalde_Shilnov,Gorbar}. The study of
the combined influence of curvature and temperature has been
performed in~\cite{Inagaki_Ishikawa}.  Recently, the dynamical chiral
symmetry breaking and its restoration for a uniformly accelerated
observer  due to the thermalization effect of acceleration was
studied in~\cite{ohsaku2} at zero chemical potential.  Further
investigations of the influence of the Unruh temperature on the phase
transitions in dense quark matter with a finite chemical potential,
and especially on the restoration of the broken color symmetry in CSC
were made in \cite{qqrindler}.

One of the widely used methods of accounting for gravitation is
based on the expansion of the fermion propagator in powers of small
curvature~$R$ \cite{Bunch_Parker,Parker_Toms}.
For instance, in \cite{kim_klim}, the three-dimensional Gross-Neveu
model in a spacetime with a weakly curved two-dimensional surface
was investigated, using an effective potential at finite curvature
and nonzero chemical potential. In paper~\cite{Goyal_Dahiya}, this 
weak curvature expansion was used in considering the D$\chi$SB at
non-vanishing temperature and chemical potential. It should, however,
be mentioned that near the phase transition point, one cannot
consider the critical curvature $R_c$ to be small and therefore 
the weak curvature expansion method can not be applied. Hence, in the
region near the critical regime different nonperturbative  methods or 
exact solutions with finite values of $R$ should be used. This kind
of solution with consideration for the chemical potential and
temperature in the gravitational background of a static Einstein
universe has been considered in~\cite{Huang_Hao_Zhuang}. There it was
demonstrated  that chiral symmetry is restored at large values of the
space curvature. Analogous studies of diquark condensation and the
related color symmetry breaking under the influence of a
gravitational field have been performed recently in the model of a
static Einstein Universe  in \cite{etz}. Recall that this model is
widely discussed in literature either as a solution of the Einstein
equations with a given cosmological constant and a nonvanishing
energy-momentum tensor of an ideal fluid as a source, or as an
initial state in inflationary cosmology with a scalar field, and the
cosmological constant as its vacuum energy (see, for instance,
\cite{ellis}). Moreover, the Einstein universe and other suitably
generalized compact curved spacetimes were extensively employed for
studying the phenomenon of Bose-Einstein condensation (see, e.g.
\cite{smith} and references therein). As is well known, one of the
possible models for the expanding universe is the closed Friedmann
model. Since the formation of quark condensates is expected to take
place considerably faster than the expansion rate of the Universe,
its radius can be considered in our calculations as constant. In this
sense, the chosen model of a static Einstein universe can be
considered as a simple cosmological model for a space with positive
curvature. On the other hand, the same form of the metric does also
hold for the interior of a (collapsing) spherically symmetric star,
thereby admitting also a non-cosmological interpretation of the
chosen model.

As was mentioned above, in quark matter there may arise another
remarkable phenomenon, pion condensation. Up to now little
information about the properties of pion condensation 
in the presence of external fields has been obtained. In particular,
there arises the interesting question, what influence gravitational 
fields can produce on pion condensation. The main aim of the present
paper is just to give a detailed investigation of this issue. 

In order to be able to consider the effects of gravitation on pion
condensation in a rigorous nonperturbative way, we will investigate
here the  phase structure of isotopically asymmetric quark matter in
the framework of an extended NJL model with dynamical  
breaking of chiral and flavor symmetries in the static Einstein
universe. In particular, our calculations are performed in
the most simple case, restricting ourselves, for simplicity, to the
flavor $SU(2)$ group. By using the mean field approximation, we will
then derive an analytical expression for the thermodynamic
potential of the NJL model that will enable us to calculate the
chiral and pion condensates of quarks moving in a space with constant
positive curvature. On this basis, we study, in particular, the phase
portraits of the model. Our paper is organized as follows. Section
II, III contain the Nambu-Jona-Lasinio model in curved spacetime and
the general formalism for the derivation of the thermodynamical
potential. Numerical calculations and conclusions are given
in sections IV, V. Finally, some estimates of the role of quantum
fluctuations in the case of a closed universe are given in the
appendix.

\section{Nambu -- Jona-Lasinio model in curved spacetime}

Suppose that dense, isotopically asymmetric quark matter (in this
case the densities of $u$ and $d$ quarks are different) in curved
spacetime is described by an extended NJL model with the following
action:
\begin{eqnarray}
&&  S=\int d^4x\sqrt{-g}~{\mathcal L}_q, \label{1}
\end{eqnarray}
and the Lagrangian
\begin{eqnarray}
&& {\mathcal L}_q =\bar q\Big [i\gamma^\nu\nabla_\nu-
m+\mu\gamma^0+\delta\mu\tau_3\gamma^0\Big ]q+ G\Big [(\bar qq)^2+
(\bar qi\gamma^5\vec\tau q)^2\Big ],  \label{2}
\end{eqnarray}
where the quark field $q(x)\equiv q_{i\alpha}(x)$ is a flavor doublet
($i=1,2$ or $i=u,d$) and color triplet ($\alpha=1,2,3$) as
well as a four-component Dirac spinor (the summation in (\ref{2})
over flavor, color and spinor indices is implied); $\tau_i$
($i=1,2,3$) are Pauli matrices. In  4-dimensional curved spacetime
with signature $(+,-,-,-)$, the line element is written as
\[
ds^2=\eta_{\hat a\hat b}e_\mu^{\hat a}e_\nu^{\hat b}dx^\mu dx^\nu.
\]
The gamma-matrices $\gamma_{\mu}$, the metric $g_{\mu\nu}$ and the
vierbein $e^{\mu}_{\hat{a}}$, as well as the definitions of the
spinor covariant derivative $\nabla_{\nu}$ and
spin connection $\omega^{\hat{a}\hat{b}}_{\nu}$ are given by the
following relations~\cite{Parker_Toms}:
\begin{eqnarray}
& & \{\gamma_{\mu}(x),\gamma_{\nu}(x)\}=2g_{\mu\nu}(x), \quad
  \{\gamma_{\hat{a}},\gamma_{\hat{b}}\}=2\eta_{\hat{a}\hat{b}}, \quad
  \eta_{\hat{a}\hat{b}}={\rm diag}(1,-1,-1,-1), \nonumber \\
& & g_{\mu\nu}g^{\nu\rho}=\delta^{\rho}_{\mu}, \quad
  g^{\mu\nu}(x)=e^{\mu}_{\hat{a}}(x)e^{\nu \hat{a}}(x), \quad
  \gamma_{\mu}(x)=e^{\hat{a}}_{\mu}(x)\gamma_{\hat{a}}. \label{2_2}\\
 & &
  \nabla_{\mu}=\partial_{\mu}+\Gamma_{\mu},\quad\Gamma_{\mu}=\frac12\
  \omega^{\hat a \hat b}_{\mu}\sigma_{\hat a \hat
  b},
  \quad
   \sigma_{\hat{a}\hat{b}}=\frac{{1}}{4}[\gamma_{\hat{a}},
   \gamma_{\hat{b}}],
 \nonumber \\
& &
\omega^{\hat{a}\hat{b}}_{\mu}=\frac{1}{2}e^{\hat{a}\lambda}
e^{\hat{b}\rho}[C_{\lambda\rho\mu}-C_{\rho\lambda\mu}-C_{\mu\lambda
\rho}],
 \quad C_{\lambda\rho\mu}=
e^{\hat{a}}_{\lambda}\partial_{[\rho}e_{\mu]\hat{a}}.
\label{3_3}
\end{eqnarray}
Here, the index $\hat{a}$ refers to the flat tangent space defined by
the vierbein at spacetime point $x$, and the $\gamma^{\hat{a}}
(\hat a=0,1,2,3)$ are the usual Dirac gamma-matrices  of Minkowski
spacetime. Moreover $\gamma_5$,  is defined, as usual (see, e.g.,
\cite{Parker_Toms}), i.e. to be the same as in
flat spacetime and thus independent of spacetime variables.

In order to describe the quark composition of matter we introduced
 $\mu=\mu_B/3$ in (\ref{2}) as the quark number chemical potential.
Since the generator $I_3$ of the third component of isospin is equal
to $\tau_3/2$, the quantity $\delta\mu$ in (\ref{2}) is half
the isospin chemical potential, $\delta\mu=\mu_I/2$. If the bare
quark mass $m$ is equal to zero, then at $\delta\mu =0$,
apart from the trivial color SU($3$) symmetry, the Lagrangian
(\ref{2}) is invariant under the chiral group $SU_L(2)\times SU_R(2)$
transformations. However, at $\delta\mu \ne 0$ this symmetry is
reduced to the subgroup $U_{I_3L}(1)\times U_{I_3R}(1)$ (here and
above the subscripts $L,R$ imply that the corresponding group acts
only on the left, right handed spinors, respectively). It is
convenient to present this symmetry as $U_{I_3}(1)\times U_{AI_3}
(1)$, where $U_{I_3}(1)$ and $ U_{AI_3}(1)$ are the isospin and axial
isospin subgroup of the chiral $SU_L(2)\times SU_R(2)$ group. Quarks
are tranformed under these subgroups in the following way 
\begin{eqnarray}
U_{I_3}(1):~~~~
q\to\exp
(\mathrm{i}\alpha\tau_3) q,~~~~~~~~~U_{AI_3}(1):~~~q\to\exp
(\mathrm{i} \alpha^\prime\gamma^5\tau_3) q, \label{2010} 
\end{eqnarray}
where $\alpha$, $\alpha^\prime$ are independent group parameters.
At nonzero bare quark mass, $m\ne 0$, and nonzero isotopic chemical
potential, i.e. $\delta\mu\ne 0$, the Lagrangian (\ref{2}) is still
invariant under the isospin subgroup $U_{I_3}(1)$, but  invariance
with respect to $ U_{AI_3}(1)$ does no longer hold in this case.

The linearized version of Lagrangian (\ref{2}), that contains
collective bosonic fields $\sigma (x)$ and $\pi_k (x)$ $(k=1,2,3)$, 
has the following form
\begin{eqnarray}
\tilde {\mathcal L}\ds &=&\bar q\Big [i\gamma^\nu \nabla_\nu
+\mu\gamma^0
+ \delta\mu\tau_3\gamma^0-\sigma - m -i\gamma^5\pi_k\tau_k\Big ]q
 -\frac{1}{4G}\Big [\sigma\sigma+\pi_k\pi_k\Big ].
\label{3}
\end{eqnarray}
From the Lagrangian (\ref{3}), one can find the equations of motion
for the  bosonic fields,
\begin{eqnarray}
\sigma(x)=-2G(\bar qq);~~~\pi_k (x)=-2G(\bar q
\mathrm{i}\gamma^5\tau_k q).
\label{200}
\end{eqnarray}
It is clear from these relations that under
isospin $U_{I_3}(1)$ and axial isospin $U_{AI_3}(1)$ transformations
(\ref{2010}) the bosonic fields (\ref{200}) are changed in the
following way:
\begin{eqnarray}
U_{I_3}(1):&&\sigma\to\sigma;~~\pi_3\to\pi_3;~~\pi_1\to\cos
(2\alpha)\pi_1+\sin (2\alpha)\pi_2;~~\pi_2\to\cos (2\alpha)\pi_2-\sin
(2\alpha)\pi_1,\nonumber\\
U_{AI_3}(1):&&\pi_1\to\pi_1;~~\pi_2\to\pi_2;~~\sigma\to\cos
(2\alpha^\prime)\sigma+\sin (2\alpha^\prime)\pi_3;~~\pi_3\to\cos
(2\alpha^\prime)\pi_3-\sin (2\alpha^\prime)\sigma.
\label{201}
\end{eqnarray}

In the fermion one-loop (mean field) approximation, the effective
action for the boson fields is expressed through the path integral
over quark fields:
$$
\exp(i {\cal S}_{\rm {eff}}(\sigma,\pi_k))=
  N'\int[d\bar q][dq]\exp\Bigl(i\int d^4 x\sqrt{-g }~\tilde
  {\mathcal L}\,\Bigr),
$$
where
\begin{equation}
{\cal S}_{\rm {eff}}
(\sigma,\pi_k)
=-\int d^4x\sqrt{-g }\left [\frac{\sigma^2+\pi^2_k}{4G}
\right ]+\tilde {\cal S}_{\rm {eff}},
\label{4}
\end{equation}
$N'$ is a normalization constant. The quark contribution to the
effective action, i.e. the term
$\tilde {\cal S}_{\rm {eff}}$ in (\ref{4}), is given by:
\begin{equation}
\exp(i\tilde {\cal S}_{\rm {eff}})=N'\int [d\bar
q][dq]\exp\Bigl(i\int d^4x\sqrt{-g }~\bar
q{\cal D}q\,\Bigr)=\mbox {det}~{\cal D}.
\label{5}
\end{equation}
In (\ref{5}), we have used the following notation
\begin{equation}
{\cal D}=i\gamma^\nu \nabla_\nu +\mu\gamma^0
+ \delta\mu\tau_3\gamma^0-\sigma - m -i\gamma^5\pi_k\tau_k.
\label{6}
\end{equation}
Clearly, ${\cal D}$ is an  operator in the coordinate, spinor
and flavor spaces. Apart from this, it is proportional to the unit
operator $1\hspace{-1.2mm}1_c$ in the color space. Thus, it can be
presented in the flavor$\otimes$color space  in the following
matrix form:
\begin{equation}
{\cal D}=\left (\begin{array}{cc}
A~, & B\\
 \bar B~, & \bar A
\end{array}\right )_f\otimes 1\hspace{-1.2mm}1_c\equiv D\otimes
1\hspace{-1.2mm}1_c,
\label{7}
\end{equation}
where operators $A,\bar A,B,\bar B$ act in the coordinate and spinor
spaces only, and
\begin{eqnarray}
A=i\gamma^\nu \nabla_\nu +(\mu+\delta\mu)\gamma^0
-\sigma - m -i\gamma^5\pi_3,&&~~~
B=i\gamma^5(\pi_1-i\pi_2),\nonumber\\
\bar A=i\gamma^\nu \nabla_\nu +(\mu-\delta\mu)\gamma^0
-\sigma - m +i\gamma^5\pi_3, &&~~~\bar
B=i\gamma^5(\pi_1+i\pi_2).
\label{8}
\end{eqnarray}
Due to the trivial color structure, it follows from (\ref{7}) that
\begin{equation}
\mbox {det}~{\cal D}=\left (\mbox {det}~D\right )^3.
\label{9}
\end{equation}
Now, suppose that $\pi_3=\pi_2=0$ and $\sigma,\pi_1$ are quantities
independent of coordinates
\footnote{To justify this assumption, let
us consider for simplicity the case $m=0$. If the fields
$\sigma,\pi_k$  do not depend on coordinates, then the effective
action is a function of the invariants $\sigma^2+\pi_3^2$ and
$\pi_1^2+\pi_2^2$ only (this fact is due to the symmetry of the model
with respect to the transformations (\ref{201})). Therefore, without
loss of generality one can put $\pi_3=\pi_2=0$.}.
In what follows, we shall denote the pion condensate $\pi_1$ by
$\Delta$ for convenience. Then, using the general formula
\begin{eqnarray}
\det\left
(\begin{array}{cc}
A~, & B\\
\bar B~, & \bar A
\end{array}\right )=\det [-\bar BB+\bar B
A\bar B^{-1}\bar A]=\det
[\bar AA-\bar AB\bar A^{-1}\bar B],
\label{10}
\end{eqnarray}
we obtain
\begin{equation}
{\cal S}_{\rm {eff}} (\sigma,\Delta) =-\int d^4x\sqrt{-g }\left
[\frac{\sigma^2+\Delta^2}{4G} \right ]-3i\ln {\rm det}D \equiv
-\Omega(\sigma,\Delta;\mu,\delta\mu)\int d^4x\sqrt{-g}, \label{11}
\end{equation}
where we have introduced the thermodynamic potential
$\Omega(\sigma,\Delta;\mu,\delta\mu)$ of the system at zero
temperature and
\begin{equation}
{\rm det}D={\rm det}\Big\{\Delta^2+[-i\gamma^\nu \nabla_\nu
-(\mu+\delta\mu)\gamma^0 -\sigma - m][i\gamma^\nu \nabla_\nu
+(\mu-\delta\mu)\gamma^0 -\sigma - m ]\Big\}. \label{12}
\end{equation}

\section{Thermodynamic potential}\label{EU}

\subsection{General formalism}

The line element in the static Einstein universe is defined by the
following relation:
\begin{equation}
ds^2= g_{\mu\nu}(x)dx^\mu dx^\nu\equiv
dt^2-a^2(d\chi^2+\sin^2\chi(d\theta^2+ \sin^2\theta d\varphi^2)),
\label{17}
\end{equation}
where $a$ is the radius of the Einstein universe (this quantity is
related to the scalar curvature, $R=6/a^2$);
$-\infty <t<\infty$, $0\leq\chi\leq\pi$, $0\leq\theta\leq\pi$,
$0\leq\varphi\leq 2\pi $. Clearly, $\gamma^0(x)$ in this case
anticommutes with all $\gamma^k(x)$ and commutes with all
$\nabla_\nu$, where $\nu=0,1,2,3$ and $k=1,2,3$. Moreover,
$\nabla_0=\partial_0$. So, starting from (\ref{12}), we have
($\nu=0,1,2,3$; $k,l=1,2,3$):
\begin{eqnarray}
{\rm det}D&=&{\rm det}\Big\{\Delta^2+(\gamma^\nu \nabla_\nu)^2
-i(\mu+\delta\mu)\gamma^0\gamma^\nu \nabla_\nu
-i(\mu-\delta\mu)\gamma^\nu \nabla_\nu\gamma^0
-\mu^2+\delta\mu^2+2\delta\mu(\sigma+m)\gamma^0+(\sigma+m)^2\Big\}
\nonumber\\
&=&{\rm
det}\Big\{\Delta^2+\partial_0^2-2i\mu\partial_0-\mu^2+\gamma^k
\nabla_k\gamma^l\nabla_l+(\sigma+m)^2-2\delta\mu[i\gamma^0\gamma^k
\nabla_k-(\sigma+m)\gamma^0] +\delta\mu^2\Big\}. \label{18}
\end{eqnarray}
Let $\hat P_0\equiv i\partial_0$, $\hat{\cal H}=-i\gamma^0\gamma^k
\nabla_k+(\sigma+m)\gamma^0$. It is evident that $\hat{\cal
H}\hat{\cal
H}=\gamma^k\nabla_k\gamma^l\nabla_l+(\sigma+m)^2$. Hence, we have
from (\ref{18})
\begin{eqnarray}
&&{\rm det}D={\rm det}\hat O,~~~~~\mbox{where}~~~~~ \hat
O=\Delta^2-(\hat P_0+\mu)^2+(\hat{\cal H}-\delta\mu)^2. \label{19}
\end{eqnarray}

Evidently, $\hat{\cal H}$ is an operator in the Hilbert space of
functions, depending on the space coordinates $\vec x$. As it is
well-known (see, e.g. \cite{cam,w}), each of the eigenvalues $\pm
E_l$ of this operator is $d_l$-fold degenerate,
\begin{eqnarray}
&&E_l=\sqrt{\omega_l^2+(m+\sigma)^2},~~~
\omega_l=\frac1a(l+\frac32),
~~~d_l=2(l+1)(l+2),~~~l=0,1,2\ldots. \label{190}
\end{eqnarray}
Thus, one can write
\begin{eqnarray}
&&\hat{\cal H}\psi_{l\alpha\eta}(\vec x)=\eta
E_l\psi_{l\alpha\eta}(\vec x),~~~~~\int d^3\vec
x\sqrt{-g}\psi_{l\alpha\eta}(\vec x)\psi_{l'\alpha '\eta
'}(\vec x)=\delta_{ll'}\delta_{\alpha\alpha '}\delta_{\eta\eta '},
\label{191}
\end{eqnarray}
where the eigenfunctions $\psi_{l\alpha\eta}(\vec x)$ ($\alpha
=1,...,d_l; \eta=\pm 1$) of the operator $\hat{\cal H}$ are also
known (see, e.g., \cite{cam,w}), and $g={\rm det}~g_{\mu\nu}=-{\rm
det}~g_{ij}(\vec x)$ (see (\ref{17})). Now, let us choose in
the Hilbert space of functions a basis of the form $\Psi_{l\alpha\eta
p_0}(t,\vec x)\equiv$ $e^{-ip_0t}\psi_{l\alpha\eta}(\vec x)$,  where
$-\infty<p_0<\infty$. Since each element $\Psi_{l\alpha\eta
p_0}(t,\vec x)$ of this basis is an eigenfunction both of $\hat P_0$
and $\hat{\cal H}$, one can easily conclude that the operator $\hat
O$ from (\ref{19}) is diagonal in this basis, i.e. each
$\Psi_{l\alpha\eta p_0}(t,\vec x)$ is an eigenfunction of $\hat O$.
The corresponding eigenvalues ${\cal E}_{l\eta p_0}$ of $\hat O$ have
the following form:
\begin{eqnarray}
&&{\cal E}_{l\eta p_0}=\Delta^2-(p_0+\mu)^2+(\eta E_l+\delta\mu)^2.
\label{192}
\end{eqnarray}
It is clear from (\ref{192}) that eigenvalues ${\cal E}_{l\eta
p_0}$ of the operator $\hat O$ do not depend on the quantum number
$\alpha=1,...,d_l$, being $d_l$-fold degenerate. Taking into account
in (\ref{11}) the relations (\ref{19}) and Det$\hat O=\exp
($Tr$\ln\hat O)$, as well as the results of Appendix \ref{ApB}, 
where the quantity Tr$\ln\hat O$ is calculated (see (\ref{B11})), one
finds
\begin{equation}
{\cal S}_{\rm {eff}}(\sigma,\Delta)+\int d^4x\sqrt{-g}\left
[\frac{\sigma^2+\Delta^2}{4G}\right ]=-3i{\rm Tr}\ln\hat O=-3i {\cal
T}\sum_{l\eta}\int\frac{dp_0}{2\pi}d_l\ln\big
[\Delta^2-(p_0+\mu)^2+(\eta E_l+\delta\mu)^2\big ]. \label{193}
\end{equation}
Here $\int d^4x\sqrt{-g}\equiv {\cal T}{\cal V}$, where ${\cal
T}=\int dt$ stands for an infinite time interval and ${\cal V}=\int
d^3\vec x\sqrt{-g}=2\pi^2a^3$ is the space volume of the Einstein
universe (the last relations are due to the fact that $g=-{\rm
  det}~g_{ij}(\vec x)$ depends only on $\vec x$).
Now, using the definition (\ref{11}) of the thermodynamic potential
(TDP) and summing in (\ref{193}) over $\eta=\pm 1$, we have for the
zero temperature case 
\begin{equation}
\Omega(\sigma,\Delta;\mu,\delta\mu)=\frac{\sigma^2+\Delta^2}{4G}
+\frac{3i}{{\cal V}}
\sum_{l=0}^\infty\int\frac{dp_0}{2\pi}d_l\big\{\ln\big
[\Delta^2-(p_0+\mu)^2+(E_l-\delta\mu)^2\big ]+ \ln\big
[\Delta^2-(p_0+\mu)^2+(E_l+\delta\mu)^2\big ]\big\}. \label{194}
\end{equation}

To find the TDP $\Omega(\sigma,\Delta;\mu,\delta\mu,T)$ in the case
of nonzero temperature $T$, one should use the imaginary time
technique, where, after summation over Matsubara frequencies (see,
e.g., \cite{etz,k1}), the following expression can be found
\begin{eqnarray}
\Omega(\sigma,\Delta;\mu,\delta\mu,T)&=&\frac{\sigma^2+\Delta^2}{4G}
-\frac{3}{{\cal V}}\sum_{l=0}^\infty
d_l\big\{E_l^{(+)}+E_l^{(-)}\big\} -\frac{3T}{{\cal V}}
\sum_{l=0}^\infty d_l\big\{\ln\big [1+e^{-\beta (E_l^{(+)}+\mu)}\big
]+ \ln\big [1+e^{-\beta (E_l^{(+)}-\mu)}\big ]\big\}
\nonumber\\&-&\frac{3T}{{\cal V}} \sum_{l=0}^\infty d_l\big\{\ln\big
[1+e^{-\beta (E_l^{(-)}+\mu)}\big ]+ \ln\big [1+e^{-\beta
(E_l^{(-)}-\mu)}\big ]\big\} \label{195}
\end{eqnarray}
with $E_l^{(\pm)}=\sqrt{(E_l\pm\delta\mu)^2+\Delta^2}$ and $\beta
=1/T$. It is clear that $\Omega(\sigma,\Delta;\mu,\delta\mu,T)$ is
an even function with respect to each of the transformations $\mu\to
-\mu$ or $\delta\mu\to -\delta\mu$. Thus, one can deal only with
non-negative values of the chemical potentials, $\mu\ge 0$,
$\delta\mu\ge 0$. 

From this moment on, we will consider only the case of nonzero
isospin chemical potential $\delta \mu \ne 0$, whereas the baryon
chemical potential is set equal to zero, $\mu=0$, since its presence
is not of principle importance for us. So, at $\mu=0$ two particular
cases can be investigated on the basis of the TDP (\ref{195}).

First, let us choose $T=0$ and $\mu=0$, but $\delta\mu\ne 0$. Then we
obtain from (\ref{195}) the expression:
\begin{eqnarray}
\Omega(\sigma,\Delta;\mu=0,\delta\mu,T=0)\equiv\Omega(\sigma,\Delta;
\delta\mu) =\frac{\sigma^2+\Delta^2}{4G} -\frac{3}{{\cal
V}}\sum_{l=0}^\infty d_l\big\{E_l^{(+)}+E_l^{(-)}\big\}. \label{197}
\end{eqnarray}
Secondly, at $T\ne 0$, $\mu =0$, $\delta\mu\ne 0$ we obtain:
\begin{eqnarray}
&&\Omega(\sigma,\Delta;\mu=0,\delta\mu,T)\equiv\Omega(\sigma,\Delta;
\delta\mu,T)=\frac{\sigma^2+\Delta^2}{4G}-\frac{3}{{\cal
V}}\sum_{l=0}^\infty d_l\big\{E_l^{(+)}+E_l^{(-)}\big\}
\nonumber\\&&~~~~~~~~~~~~~~~~~~~~~~~~~ -\frac{6T}{{\cal V}}
\sum_{l=0}^\infty d_l\big\{\ln\big [1+e^{-\beta E_l^{(+)}}\big ]+
\ln\big [1+e^{-\beta E_l^{(-)}}\big ]\big\}. \label{198}
\end{eqnarray}

Next, let us consider the limit of zero curvature or infinitely large
radius of the universe. It is clear that the metric (\ref{17}) never
coincides with that of the flat Minkowsky spacetime because these
two spacetimes have different topologies. However, in the limit
$a\to\infty$ and $R\to 0$ one can obtain from (\ref{198}) the usual
expression for the TDP in flat spacetime (see, e.g., \cite{[17]}) 
by the following substitution:
$$
\frac l a \to k,\quad \omega_l\to k, \quad d_l=2(l+1)(l+2) \to
2k^2a^2,\quad \sum_l \to \int dl=a\int dk
$$
In this case, the TDP looks as follows:
\begin{eqnarray}
&&\Omega(\sigma,\Delta;
\delta\mu,T)=\frac{\sigma^2+\Delta^2}{4G}-6\int
\frac{d^3k}{(2\pi)^3} \left\{E_k^{(+)}+E_k^{(-)} + 2T
\ln[1+e^{-\beta E_k^{(+)}}]+2T \ln[1+e^{-\beta E_k^{(-)}}]\right\},
 \label{199}
\end{eqnarray}
where $E_k^{(\pm)}=\sqrt{(E_k\pm \delta\mu)^2+\Delta^2}$.

It should be noted that one more particular case, when $T\ne 0$,
$\mu\ne 0$, but $\delta\mu =0$, can easily be reduced to the
investigation of the one-flavored NJL model at $T\ne 0$, $\mu\ne 0$
in the Einstein universe \cite{Huang_Hao_Zhuang}, and hence we shall
not consider it here.

\subsection{Regularization}

First of all, in order to normalize the TDP, we should subtract a
corresponding constant from it, such that $\Omega(\sigma=0,\Delta=0)
=0$. The thermodynamic potential, normalized in this way, is still
divergent at large $l$, and hence, we should introduce a (soft)
cutoff in the summation over $l$ by means of the multiplier
$e^{-\omega_l/\Lambda}$ \cite{Huang_Hao_Zhuang,etz}, where
$\Lambda$ is the cutoff parameter
\footnote{In flat spacetime, the cutoff constant $\Lambda$ can be
specified according to the experimental results. However, in the case
of a curved spacetime, in order to fix the cutoff $ \Lambda$, we need
new theoretical/experimental inputs for chiral QCD in (strong)
gravitational background fields, concerning, for instance, an
effective gluon mass, known values of the quark condensate or even
experimentally measured characteristics of a pion. 
Due to the lack of experimental knowledge, our aim is here
to perform only a qualitative study of gravitational 
effects on the quark and pion condensates by investigating the 
respective phase portraits of the NJL model.  For this reason it is
convenient  to scale the thermodynamic potential and all relevant
quantities like condensates, curvature, chemical potential and
temperature by the cutoff $\Lambda$.}.

For convenience, we shall multiply all dimensional quantities that
enter the thermodynamic potential by the corresponding power of
$\Lambda$ to make them dimensionless, i.e., $\Omega/\Lambda^4, \;
\sigma/\Lambda, \; \Delta/\Lambda, \; \Lambda^{2}G, \; \Lambda^{3}V,
\; R/\Lambda^2, \; T/\Lambda, \; \mu/\Lambda, \;
\delta\mu/\Lambda,\; \omega_l/\Lambda$, and denote them using the
same letters as before: $\Omega, \;
 \sigma, \; \Delta, \; G, \; V, \; R,
\; T, \; \mu, \; \delta\mu, \;\omega_l$. Then the regularized
thermodynamic potential is written as
\begin{eqnarray}
&&\Omega^{reg}(\sigma,\Delta;
\delta\mu,T)=\frac{\sigma^2+\Delta^2}{4G}-\frac{3}{{\cal
V}}\sum_{l=0}^\infty e^{-\omega_l}
d_l\big\{E_l^{(+)}+E_l^{(-)}\big\}
\nonumber\\&&~~~~~~~~~~~~~~~~~~~~~~~~~ -\frac{6T}{{\cal V}}
\sum_{l=0}^\infty e^{-\omega_l} d_l\big\{\ln\big [1+e^{-\beta
E_l^{(+)}}\big ]+ \ln\big [1+e^{-\beta E_l^{(-)}}\big ]\big\}.
\label{198_1}
\end{eqnarray}

In the following Section, we shall perform  a numerical calculation
of the points of the global minimum of the finite regularized
thermodynamic potential $\Omega^{\rm reg}(\sigma,\Delta)-\Omega^{\rm
reg}(0,0)$ (they should of course coincide with the minima of the
potential $\Omega^{\rm reg}(\sigma,\Delta)$), and with the use of
them, consider quark matter phase transitions in the gravitational
field of the Einstein universe.

\section{Numerical calculations}

In this section, on the basis of the thermodynamic potentials
(\ref{197})-(\ref{198}), we will study numerically phase transitions
in quark matter and consider only the case of nonzero isospin
chemical potential $\delta \mu \ne 0$, whereas the baryon chemical
potential is set equal to zero, $\mu=0$. In order to obtain the
values of condensates, one should find the global minimum point (GMP)
of the thermodynamic potentials over the variables $\sigma$, $\Delta$
from the corresponding gap equations
\[
{\partial \Omega^{\rm reg}(\sigma,\Delta)\over
\partial\sigma}=0,\,\,\,{\partial \Omega^{\rm
reg}(\sigma,\Delta)\over \partial\Delta}=0.
\]
Formally, there are four different expressions for the GMP: 
i) $(\sigma=0,\Delta =0)$, ii) $(\sigma\ne 0,\Delta =0)$, iii)
$(\sigma=0,\Delta\ne 0)$, iv) $(\sigma\ne 0,\Delta\ne 0)$. 
The first two GMPs correspond to the isotopically invariant phases of
the model, whereas the GMPs of the form iii) and iv) correspond to
the phases, in which the ground state is no more $U_{I_3}(1)$
invariant. In these phases the pion condensation phenomenon
occurs. For simplicity, we take for the numerical calculations of the
GMP the value of the coupling constant $G=1$  (in our dimensionless
choice of parameters).

\subsection{Zero temperature}

Let us first consider phase transitions at zero temperature, $T=0$
and choose the current quark mass to be equal to zero, $m=0$. The
thermodynamic potential in this case is described by formula
(\ref{197}). 

\begin{figure}[ht]
   \noindent
 \centering
 \epsfig{file=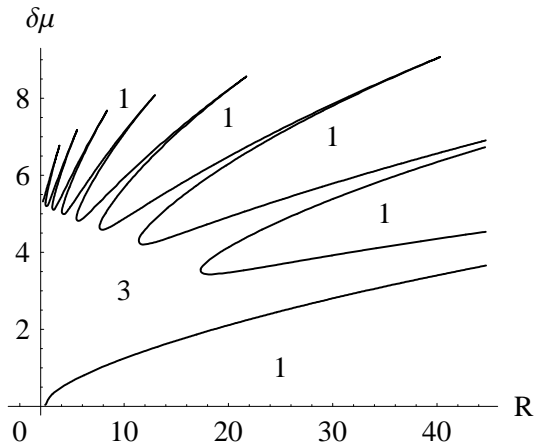,width=8cm}
  \caption{The phase portrait at zero temperature $T=0$ and $m=0$.
  Number 1 denotes the symmetric phase and 3 denotes the isospin
  symmetry breaking phase (the phase with the pion condensate 
 $\Delta\ne 0$).}
   \label{pp1}
\end{figure}

The detailed investigation of the GMP properties vs external
parameters $R$ and $\delta\mu$ results in the phase portrait shown in
Fig.~\ref{pp1}. For the points of symmetric phase 1, the GMP is at
$\sigma=0$ and $\Delta=0$. In the phase 3, the minimum is at
$\sigma=0$ and $\Delta\neq 0$, and this indicates that the isospin
$U_{I_3}(1)$ symmetry of the model is dynamically broken in this
phase. We note that at the same time the chiral $U_{AI_3}(1)$
symmetry remains unbroken in this phase, since in the GMP we have
$\sigma=0$, as it should be in the case of zero
current quark mass, when  $\Delta\neq 0$.

\begin{figure}[ht]
   \noindent
 \centering
 $
 \begin{array}{cc}
 \epsfig{file=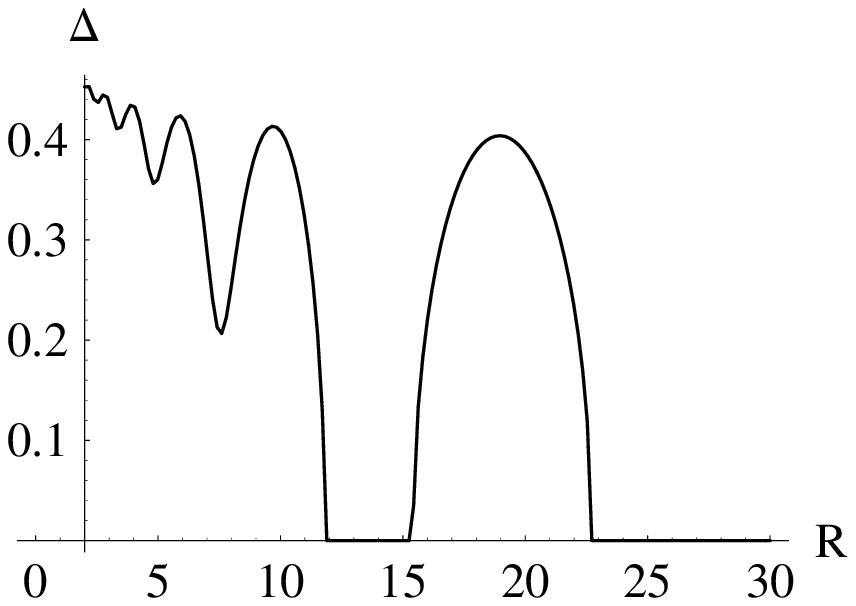,width=8cm} &
 \epsfig{file=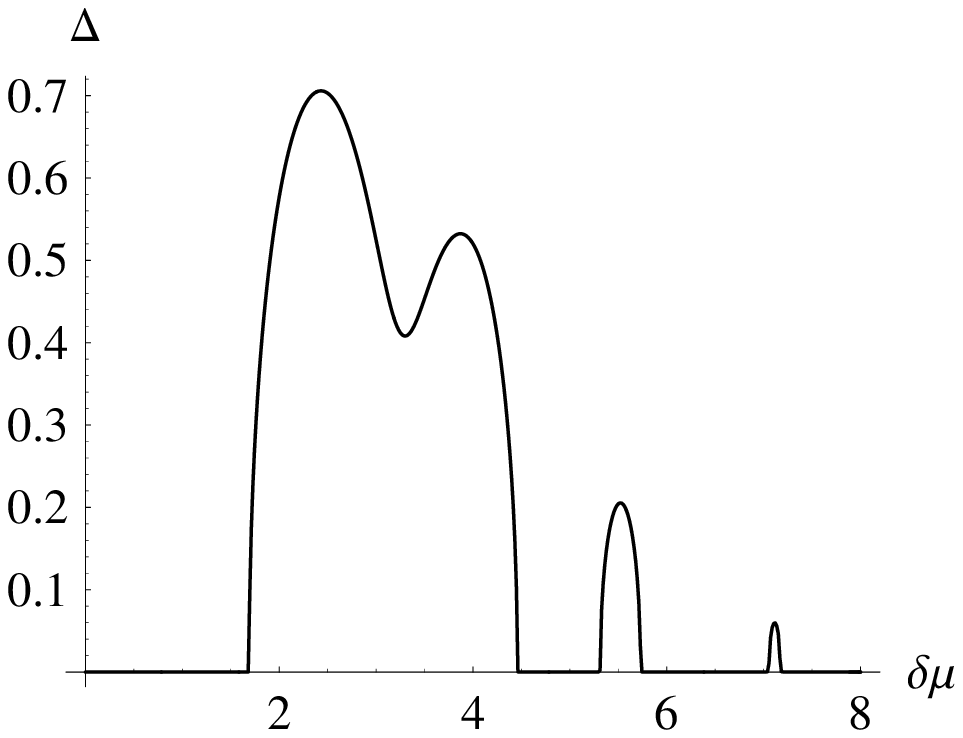,width=8cm}
 \end{array}
 $
\caption{The behaviour of the pion condensate $\Delta$. 
Left picture: $T=0, \delta\mu=4.5$. 
Right picture: $R=15, T=0$.}
\label{cond23}
\end{figure}

As one can see from Fig.~\ref{pp1}, the 
critical curve, which separates the phases 1 and 3, 
has an oscillating character. This phenomenon is 
explained by the discreteness of the fermion energy levels
(\ref{190}) in compact space. Moreover, as it is clearly seen from
the curves in \ref{cond23}, the pion condensate $\Delta$ vs $R$ also
oscillates in the phase 3. This effect resembles the van Alphen-de
Haas oscillations of different physical quantities in the magnetic
field $H$, where fermion levels are also discrete (the Landau levels)
\cite{and} (see also \cite{aand}, where a similar influence of a
magnetic field on the oscillation behavior of the Compton 
scattering and photoproduction cross-sections  was demonstrated).
Indeed, the corresponding magnetic oscillations of the critical curve
in the $\mu$-$H$ phase portrait of dense cold quark matter with
four-fermion interactions were found in papers \cite{oscil}. There, 
the existence of the standard van Alphen-de Haas magnetic
oscillations of some thermodynamical quantities, including
magnetization, pressure and particle density  of cold dense quark
matter was also demonstrated. 

The behavior of the pion condensate $\Delta$ as a function of $R$
at fixed $\delta\mu$ and as a function of $\delta\mu$ at fixed $R$
is shown in Fig.~\ref{cond23} (left  and right pictures
respectively). 

\begin{figure}[ht]
   \noindent
 \centering
 \epsfig{file=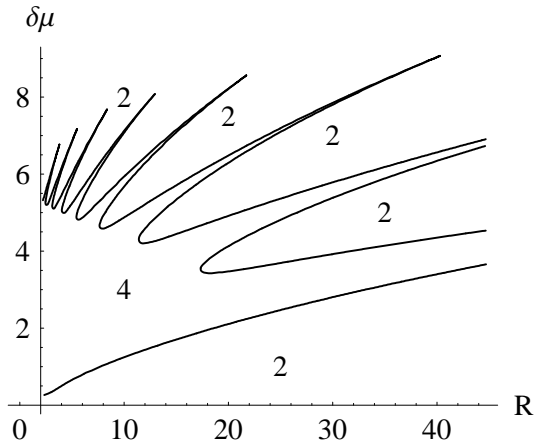,width=8cm}
  \caption{The phase portrait at zero temperature, $T=0$, and
  $m=0.01$. Number 2 denotes the chiral symmetry breaking phase with
  $\sigma \ne 0, \Delta =0$, and 4 denotes the mixed
  phase with $\sigma\ne 0$ and $\Delta \ne 0$.}
   \label{pp2}
\end{figure}

The phase portrait at finite current quark mass, $m\ne 0$,  is
depicted in Fig.~\ref{pp2}. In phase 2 the chiral symmetry is now
broken due to a finite value of the current quark mass,  and the
global minimum of TDP is at $\sigma\neq 0$ and $\Delta = 0$. In the
mixed phase 4 both condensates are nonzero, i.e. $\sigma\neq 0$ and
$\Delta\neq 0$.

\begin{figure}[ht]
   \noindent
 \centering
 $
 \begin{array}{cc}
 \epsfig{file=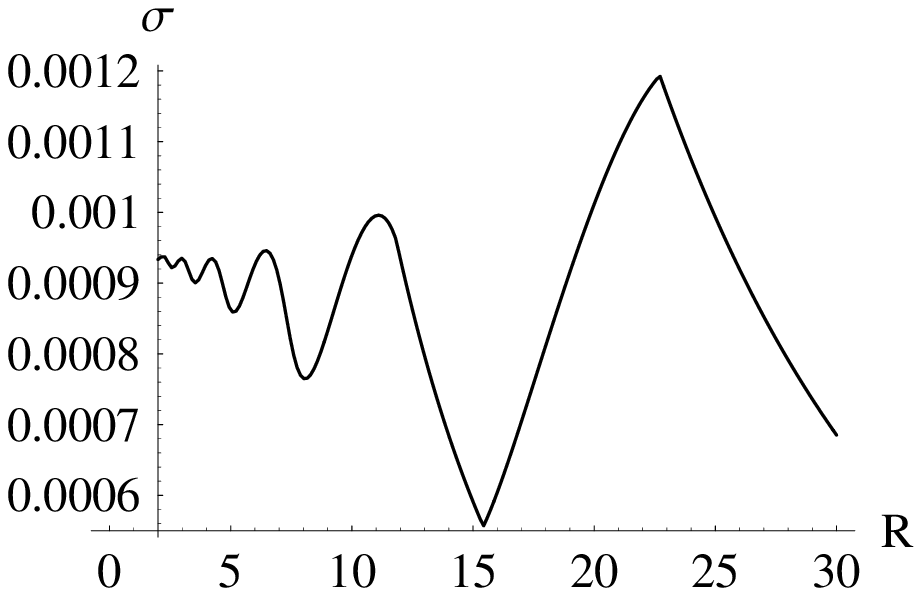,width=8cm} &
 \epsfig{file=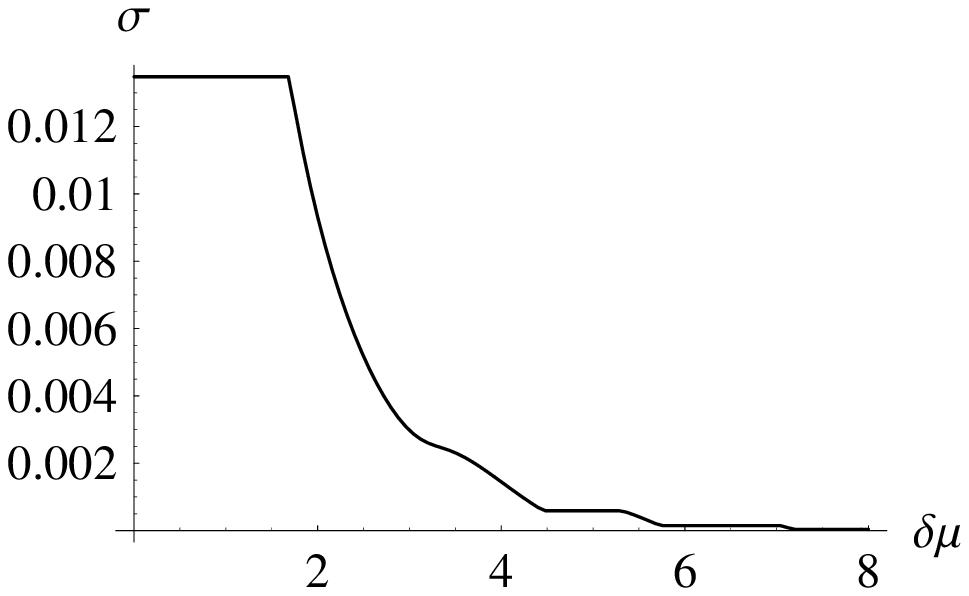,width=8cm}
 \end{array}
 $
 \caption{Condensate $\sigma$ at $\delta\mu=4.5$ (left picture) and
 $R=15$ (right picture), $m=0.01$}
   \label{cond5}
\end{figure}

The behavior of the chiral condensate $\sigma$ in the case of finite
quark mass, $m\ne 0$,  is
shown in Fig.~\ref{cond5} as a function of $R$
at $\delta\mu=4.5$ (left picture) and as a function of $\delta\mu$ at
$R=15$ (right picture).

One can see oscillations of $\sigma$  on both pictures, although
they are rather strong in the left picture for the dependence on
$R$, while they are weakly seen in the high $\delta \mu$ tail of the
curve in the right picture. One should note that the right picture
resembles, except for these oscillations, the corresponding curve in
\cite{[23]} (Fig. 1) for the flat case.

\subsection{Finite temperature}

\begin{figure}[ht]
   \noindent
 \centering
 \epsfig{file=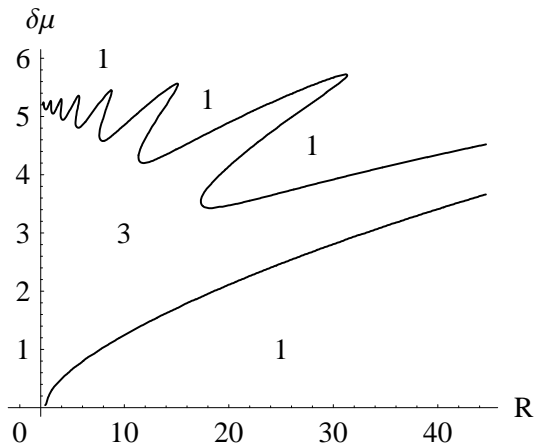,width=8cm}
  \caption{The phase portrait at $T=0.1$ and $m=0$. Number
  1 denotes the symmetric phase and 3 denotes the isospin symmetry
  breaking phase (the phase with the pion condensate 
 $\Delta\ne 0$).}
   \label{pp3}
\end{figure}

Using formula (\ref{198_1}) for the thermodynamic potential, we can
also study the influence of finite temperature on phase transitions.
The phase portrait at $T=0.1$ and zero current quark mass, $m = 0$,
is shown in Fig.~\ref{pp3} in terms of $R-\delta\mu$.
It is seen from this figure that growing temperature leads to a
smoothing of oscillations of the phase curve.

\begin{figure}[ht]
\noindent
    \centering
    $
    \begin{array}{cc}
 \epsfig{file=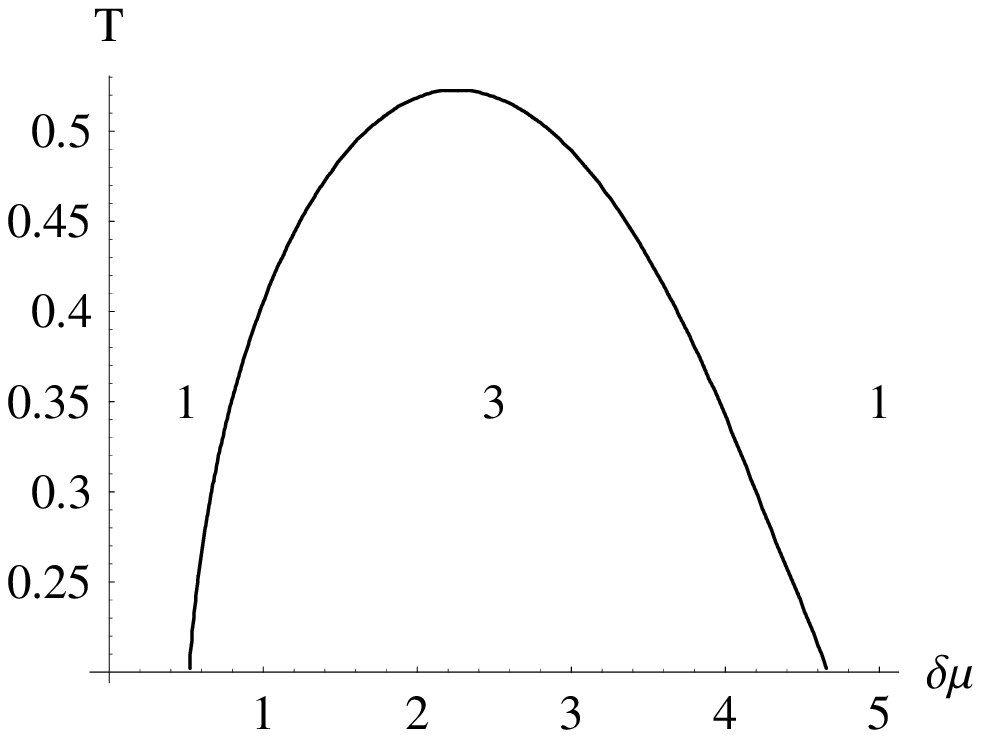,width=8cm}&
 \epsfig{file=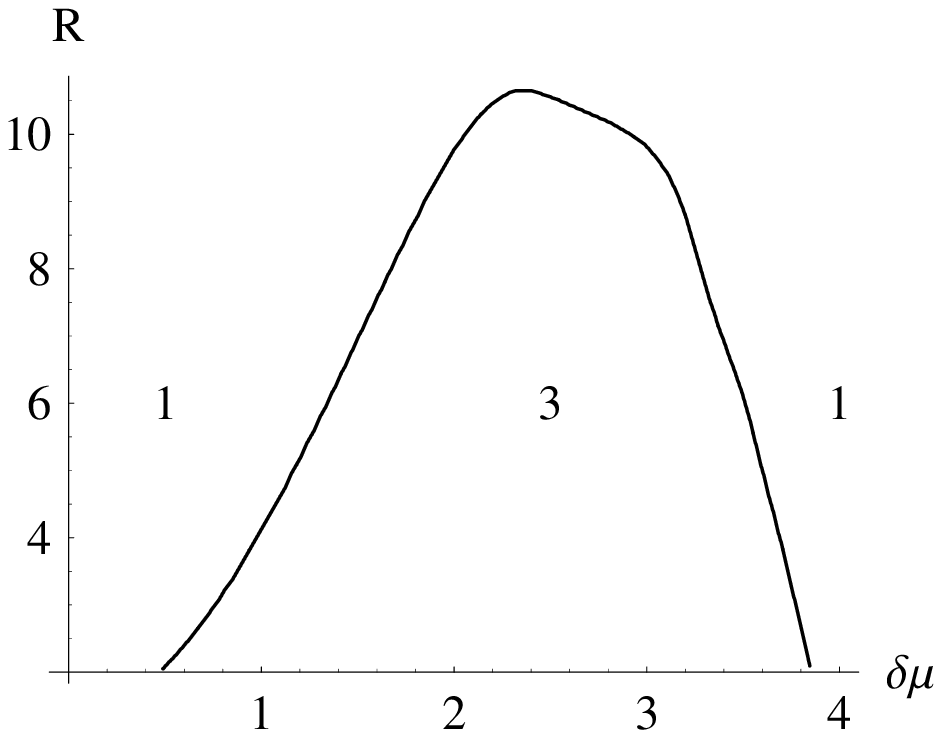,width=8cm}
\end{array}
    $
 \caption{The phase portraits at $R=4$ (left picture) and at $T=0.4$
 (right
   picture), $m=0$.} \label{pp4}
\end{figure}

For comparision, in Fig.~\ref{pp4}, the phase portraits in the 
$T-\delta \mu$ and $R-\delta\mu$ planes (the latter now at another
value of temperature $T=0.4$) are depicted. First of all, it is clear
from Fig.~\ref{pp4} that the isospin symmetry is restored due to the
vanishing of the pion condensate both at high temperature and high
curvature. The similarity between these two plots leads to the
conclusion that curvature and temperature effects play a similar role
in the restoration of isospin symmetry.

\begin{figure}[ht]
   \noindent
 \centering
 \epsfig{file=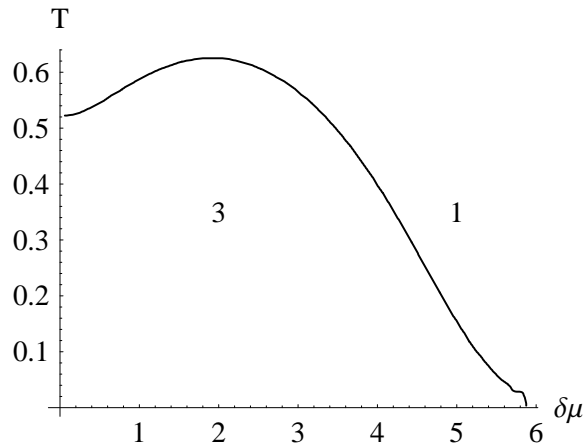,width=8cm}
  \caption{The phase portrait at $R=0$ and $m=0$.}
   \label{pp5}
\end{figure}

The phase portrait in Fig.~\ref{pp5} for $R=0$ corresponds
to the case of flat Minkowski spacetime. It looks very
similar to the one obtained, for instance, in \cite{[23]} (see upper
panel of their corresponding Fig. 12). Let us compare the phase
portraits at zero (Fig.~\ref{pp5}) and finite curvature
(Fig.~\ref{pp4}, left picture). In the first case the pion
condensation appears at arbitrary small nonzero values
of the isospin chemical potential, while in the second case the
isospin symmetry becomes dynamically broken only at some finite value
of the chemical potential $\delta\mu$. This phenomenon
may be explained by the existence of a gap in the quark spectrum
(\ref{190}), which is proportional to the inverse radius of the
Einstein universe. In this case the effect of curvature is similar
to the effect of a nonzero current quark mass (for comparison see
lower panel of Fig. 12 in \cite{[23]} and our Fig.~\ref{pp4}).

\section{Conclusions}

In the framework of an
extended Nambu--Jona-Lasinio model, we have studied the influence of
a gravitational field on pion condensation in isotopically asymmetric 
quark matter at finite temperature and isospin chemical potential.
As a particular model of a gravitational field configuration we have
taken the static Einstein universe. This particular choice enabled us
to investigate  phase transitions of the system with an exact
consideration of the role of the gravitational field in the formation
of the quark and pion condensates and thus to demonstrate its
influence on the phase portraits. In particular, we have found that
thermodynamic quantities such as quark and pion condensates as well
as the corresponding phase boundaries (critical curves) oscillate as 
functions of curvature. This oscillating behavior is smoothed out
with growing temperature. There  exists also an interesting 
similarity between the behavior of phase portraits when considered 
as functions of curvature and as functions of finite temperature
(compare Figs. 6). Moreover, we have shown that for massless quarks
and for some values of $R$ the isospin symmetry becomes in curved
spacetime dynamically broken  only at some finite value of the
chemical potential $\delta\mu$ (look at Fig. 6 (left picture) for
rather small values of $T$).  This is contrary to the case of flat
spacetime, where the pion condensation appears at arbitrary small
nonzero values of the isospin chemical potential (see Fig. 7). This
effect resembles the pion condensation but at nonzero bare quark mass
and may be explained by the presence of a gap in the energy spectrum
of quarks in the gravitational field. 

Finally, let us add some remarks on the possible role of
quantum fluctuations of the collective fields and finite size
effects in our approach. In fact, since the volume of the space
region modelled by the closed Einstein universe adopted in this paper
is limited,  finite size effects could eventually change the
character of phase transitions (see e.g. discussions in
\cite{rubakov,denardo}). Eventually, this might even lead to
particular situations, where no phase transition can occur. However,
as it is clear from physical considerations, the finite size in
itself may not practically forbid the dynamical symmetry breaking, if
the characteristic length of the region of space occupied
by the system is much greater than the Compton wavelength of the
excitation responsible for tunneling and restoration of symmetry
(see, e.g., \cite{rubakov}). It should be noted that our results,
obtained with the use of the mean field approximation in the
framework of the NJL model are consistent with corresponding 
estimates of the role of quantum fluctuations (see Appendix B).

In conclusion, we  emphasize that the results of this paper are 
evidently only of a qualitative nature. They do not allow us to find
an exact value of the critical radius, and hence, further studies
with more realistic models of gravitational fields should be
undertaken. 

\acknowledgments
Two of us (V.Zh. and A.V.T.) thank M. Mueller-Preussker for
hospitality during their stay at the Institute of Physics of
Humboldt-University, where part of this
work has been done, and also DAAD for financial support. D.E. is
grateful to R. Seiler for discussion on
the role of quantum fluctuations and finite size effects. 
This work was supported in part by the DFG-grant 436 RUS 113/477.

\appendix

\section{The trace of the operator $\hat O$ (\ref{19})}\label{ApB}

In Section \ref{EU} we have introduced two operators, $\hat P_0$ and
$\hat{\cal H}$, acting in the Hilbert space $\mathbf H$ of all
quadratically integrable functions $f(t,\vec x)$ defined on the
spacetime manifold of the Einstein universe.

Now suppose that there is an abstract Hilbert space $\mathbb
H$ of vectors $|f\rangle$. Let $\hat{\vec \mathbf x}$ and $\hat
\mathbf
t$ be the operators of the space and time coordinates,
correspondingly, defined on $\mathbb H$. Moreover, let $|t,\vec
x\rangle\equiv |x\rangle$ be the complete set, or basis, of
eigenvectors of $\hat{\vec \mathbf x}$ and $\hat
\mathbf t$, i.e. $\hat{\vec\mathbf x}|x\rangle=\vec x|x\rangle$,
$\hat
\mathbf t|x\rangle=t|x\rangle$. The set $|x\rangle$ is usualy called
the coordinate basis in $\mathbb H$. Obviously, the 
 completeness and normalization conditions for the coordinate
basis $|x\rangle$ are valid:
\begin{eqnarray}\label{B1}
&&\int d^4 x \sqrt{-g}|x\rangle\langle x|=\mathbf I,\\
&&\langle x'  |x\rangle=\frac{\delta(x-x')}{\sqrt{-g}},
\label{B2}
\end{eqnarray}
where $\mathbf I$ is the unit operator in $\mathbb H$ and $g={\rm
det}~g_{\mu\nu}$. Due to
(\ref{B1}), it is possible to expand any vector $|f\rangle\in \mathbb
H$ in terms of the basis $|x\rangle$, namely $|f\rangle=\int d^4 x
\sqrt{-g} |x\rangle\langle x|f\rangle$. The quantity  $\langle
x|f\rangle$ is called {\it $x$- (or coordinate) representation for
the vector $|f\rangle$}. Identifying $\langle x|f\rangle$ with
$f(t,\vec x)$, we see that the Hilbert space $\mathbf H$ of all
quadratically integrable functions is simply the coordinate
representation of the above introduced abstract Hilbert space
$\mathbb H$. Furthermore, if in the Hilbert space $\mathbb H$ there
is an arbitrary operator $\hat\mathbf a$, then the matrix $\hat A$,
whose matrix elements in the coordinate basis are just the quantities
$\langle x'|\hat\mathbf a|x\rangle$, is called the $x$- (or
coordinate) representation of the operator $\hat\mathbf a$.
Obviously, in order to define an operator $\hat\mathbf a$ in the
abstract Hilbert space $\mathbb H$, it is sufficient to define its
matrix $\hat A\equiv \langle x'|\hat\mathbf 
a|x\rangle$, which acts in the Hilbert space $\mathbf H$ of all
quadratically integrable functions, in the coordinate basis. 
It is clear that Tr$\hat A=$$\int 
d^4 x\sqrt{-g} \langle x|\hat\mathbf a|x\rangle$.

Now, let us consider in $\mathbb H$ the two commuting
operators $\hat\mathbf p_0$ and $\hat\mathbf h$ such that in the
$x$-representation they look like $\hat P_0\equiv \langle
x'|\hat\mathbf p_0|x\rangle$ and $\hat{\cal H}\equiv \langle x'
|\hat\mathbf h|x\rangle$, correspondingly (see 
section \ref{EU}). The operators $\hat P_0$ and $\hat{\cal H}$ have a
common set of  eigenfunctions $\Psi_{l\alpha\eta p_0}(t,\vec x)$
defined in section \ref{EU}, from which it follows that
\begin{eqnarray}
\int d^4 x \sqrt{-g}\Psi_{l\alpha\eta p_0}(t,\vec x)\Psi^*_{l'\alpha
'\eta ' p'_0}(t,\vec x)=2\pi\delta(p'_0-p_0)\delta_{ll'}
\delta_{\alpha\alpha '}\delta_{\eta\eta '}.
\label{B3}
\end{eqnarray}
The eigenfunctions $\Psi_{l\alpha\eta p_0}(t,\vec x)$ are the
coordinates of the corresponding eigenvectors $|l\alpha\eta
p_0\rangle\in \mathbb H$ of the operators $\hat\mathbf p_0$ and
$\hat\mathbf h$ (recall, $l=0,...\infty;\alpha=1,..,d_l
\equiv 2(l+1)(l+2); \eta =\pm 1; -\infty<p_0<\infty$):
\begin{eqnarray}
|l\alpha\eta p_0\rangle=\int d^4 x
\sqrt{-g}|x\rangle  \Psi_{l\alpha\eta p_0}(t,\vec x)\equiv \int d^4 x
\sqrt{-g}|x\rangle \langle x|l\alpha\eta p_0\rangle.
\label{B4}
\end{eqnarray}
(Clearly, $\hat\mathbf p_0|l\alpha\eta p_0\rangle=p_0|l\alpha\eta
p_0\rangle$ and $\hat\mathbf h|l\alpha\eta p_0\rangle=\eta
E_l|l\alpha\eta p_0\rangle$.)
It follows from (\ref{B4}) that $\Psi_{l\alpha\eta p_0}(t,\vec x)=
\langle x|l\alpha\eta p_0\rangle$. Using this relation in the
normalization condition (\ref{B3}) and then integrating there over
$x$ with the help of (\ref{B1}), we obtain
\begin{eqnarray}
\langle l'\alpha '\eta ' p'_0 |l\alpha\eta
p_0\rangle=2\pi\delta(p'_0-p_0)\delta_{ll'}\delta_{\alpha\alpha
'}\delta_{\eta\eta '}
\label{B5}
\end{eqnarray}
which is the analogue of the normalization condition (\ref{B2}).
It is possible to show that the completness condition for the basis
$|l\alpha\eta p_0\rangle$ follows from (\ref{B5}):
\begin{eqnarray}
\sum_{l\alpha\eta}\int\frac{dp_0}{2\pi}
|l\alpha\eta p_0\rangle\langle
l\alpha\eta p_0|=\mathbf I.
\label{B6}
\end{eqnarray}
Let us construct in $\mathbb H$ the following operator
\begin{eqnarray}
\hat\mathbf o=\Delta^2-(\hat\mathbf p_0+\mu)^2+(\hat\mathbf
h+\delta\mu)^2 \label{B7}
\end{eqnarray}
which is  diagonal  in the basis (\ref{B4}), i.e. each
vector $|l\alpha\eta p_0\rangle$ is its eigenvector with
corresponding eigenvalue ${\cal E}_{l\alpha\eta p_0}$ (\ref{192}). In
the coordinate representation its matrix $\langle x'|\hat\mathbf
o|x\rangle$ coincides with the operator $\hat O$ (\ref{19}). So,
\begin{eqnarray}
{\rm Tr}\hat O\equiv\int
d^4 x\sqrt{-g} \langle x|\hat\mathbf o|x\rangle=
\int d^4 x\sqrt{-g} \sum_{l\alpha\eta}\int\frac{dp_0}{2\pi}
\sum_{l'\alpha '\eta '}\int\frac{dp'_0}{2\pi} \langle x|l\alpha\eta
p_0\rangle\langle l\alpha\eta p_0|
\hat\mathbf o|l'\alpha '\eta ' p'_0\rangle\langle
l'\alpha '\eta 'p'_0|x\rangle,
\label{B8}
\end{eqnarray}
where the last equality was obtained by employing the completeness 
relation (\ref{B6}). Now, by using in this formula the
eigenvalue condition $\hat\mathbf o|l\alpha\eta p_0\rangle={\cal
E}_{l\alpha\eta p_0}|l\alpha\eta p_0\rangle$,  the normalization
condition (\ref{B5}), and, finally, by performing in the obtained
expression the integration and summation over primed indices,
we have
\begin{eqnarray}
{\rm Tr}\hat O=\int d^4 x\sqrt{-g}
\sum_{l\alpha\eta}\int\frac{dp_0}{2\pi}{\cal E}_{l\alpha\eta p_0}
\langle x|l\alpha\eta
p_0\rangle\langle l\alpha\eta p_0|x\rangle=
\int d^4 x\sqrt{-g}
\sum_{l\alpha\eta}\int\frac{dp_0}{2\pi}{\cal E}_{l\alpha\eta p_0}
\Psi_{l\alpha\eta p_0}(t,\vec x)\Psi^*_{l\alpha\eta p_0}(t,\vec x).
\label{B9}
\end{eqnarray}
Since in (\ref{B9}) the quantities $\sqrt{-g}$ and $\Psi_{l\alpha\eta
p_0}(t,\vec x)\Psi^*_{l\alpha\eta p_0}(t,\vec
x)\equiv$$\psi_{l\alpha\eta}(\vec x)\psi_{l\alpha\eta}(\vec x)$ (the
last relation is due to the notations accepted in  formula
(\ref{191}) and below) do not depend on the time coordinate, the
expression (\ref{B9}) is proportional to the infinite time interval
${\cal T}\equiv\int dt$. The remaining $\vec x$-integration in
(\ref{B9})  gives simply  unity due to the relation (\ref{191}). So
we have
\begin{eqnarray}
{\rm Tr}\hat O={\cal T} \sum_{l\alpha\eta}\int\frac{dp_0}{2\pi}{\cal
E}_{l\alpha\eta p_0}={\cal T}
\sum_{l\eta}\int\frac{dp_0}{2\pi}d_l\big [\Delta^2-(p_0+\mu)^2+(\eta
E_l+\delta\mu)^2\big ], \label{B10}
\end{eqnarray}
where the fact that each eigenvalue ${\cal E}_{l\alpha\eta
p_0}$ is $d_l$-fold degenerated is taken into account and the
notations from (\ref{190})-(\ref{192}) are used. In a similar way it
is possible to obtain the quantity Tr$\ln\hat O$:
\begin{eqnarray}
{\rm Tr}\ln\hat O={\cal T}
\sum_{l\alpha\eta}\int\frac{dp_0}{2\pi}\ln{\cal E}_{l\alpha\eta
p_0}={\cal T} \sum_{l\eta}\int\frac{dp_0}{2\pi}d_l\ln\big
[\Delta^2-(p_0+\mu)^2+(\eta E_l+\delta\mu)^2\big ]. \label{B11}
\end{eqnarray}

\section{The role of quantum fluctuations and finite size
effects}\label{ApC}

It is well known that spontaneous symmetry breaking in low
dimensional quantum field theories may become impossible due to
strong quantum fluctuations of fields. The same is also true for
systems that occupy a limited space volume. However, as it is clear
from physical considerations, the finite size in itself may not
practically forbid the spontaneous symmetry breaking, if the
characteristic length of the region of space occupied by the system
is much greater than the Compton wavelength of the excitation
responsible for tunneling and restoration of symmetry. (Indeed, one
may recall here well known physical phenomena such as the
superfluidity of Helium or superconductivity of metals that are
observed in samples of finite volume). This idea has been discussed
for some scalar field theories in the Einstein universe for
instance, in \cite{rubakov,denardo}. In this Appendix, we shall
demonstrate that, under certain conditions, dynamical symmetry
breaking in NJL-type models is indeed possible in the closed Einstein
universe. In particular, we will show that, if the radius of the
universe is large enough such that the fluctuations of quantum fields
are comparatively small, the symmetry breaking obtained in the mean
field approximation is not forbidden. 

For illustrations, let us confine to the analogous case of the chiral
condensate and consider the simplified case of the
linearized Lagrangian (\ref{3}) with $\mu=0, \delta\mu=0, m=0,
\pi_k=0 $, 
\[
\tilde {\mathcal L}\ds =\bar q(i\gamma^\nu \nabla_\nu-\sigma)q
 -\frac{1}{4G}\sigma^2, 
\]
and the corresponding partition function 
$Z={\rm e}^{i{\cal S}_{\rm {eff}}}$ with
\begin{equation}
{\cal S}_{\rm {eff}} (\sigma) =-\int d^4x\sqrt{-g
}\frac{\sigma^2}{4G}-i{\rm Tr} \ln (i\gamma^\nu \nabla_\nu
 -\sigma). 
\end{equation}
Now, supposing that $\sigma=\sigma_0+\phi$, where
$\sigma_0$ is the vacuum expectation value of the field $\sigma$ and
$\phi$ denotes its quantum fluctuation, we obtain: 
\[
\ln({\cal D}-\phi)\approx \ln {\cal D}-{\cal D}^{-1}\phi-\frac
   {1}{2}({\cal D}^{-1}\phi)({\cal D}^{-1}\phi)-\dots ,
\]
where ${\cal D}=i\gamma^\nu \nabla_\nu
-\sigma_0$. Thus $Z=Z_0Z_{\phi}$, where 
$Z_0=\exp i S_0$, $S_0=-\int d^4x\sqrt{-g} V_0$, and
\begin{equation}
V_0=\frac {\sigma_0^2}{4 G}
+i \frac {1}{\int d^4x\sqrt{-g}}{\rm Tr}\ln
{\cal D}
\label{v01}
\end{equation}
is the effective potential at $\sigma=\sigma_0$. The contribution
of quantum fluctuations  up to the $\phi^2$-term to the effective
action $S_\phi$ is given by
\[
Z_{\phi}\equiv\int d\phi \exp(iS_\phi)=\int d\phi\exp\left\{-i\int
  d^4x\sqrt{-g}\frac {1}{4G}(\phi^2+2\sigma_0\phi)-{\rm Tr} ({\cal
  D}^{-1}\phi +\frac
  {1}{2}({\cal D}^{-1}\phi)({\cal D}^{-1}\phi))\right\}. 
\]
It is evident, that in the above expansion the  term linear in $\phi$
corresponds to the so-called tadpole diagram with one external
$\phi$-line and the term quadratic in $\phi$ corresponds to the
``polarization operator'' diagram of the $\phi$ field with one
fermion loop. From (\ref{v01}) we can write the stationarity
condition and find the gap equation, $\partial V_0/\partial \sigma_0
=0$,
\begin{equation}
\frac{\sigma_0}{2G}\int d^4x\sqrt{-g}=i{\rm Tr} {\cal D}^{-1},
\label{gap}
\end{equation}
and the linear terms in $\phi$, corresponding to the tadpole diagram,
cancel out in $Z_{\phi}$. Thus, the contribution of fluctuations to 
the field action is given by 
\begin{equation}
S_\phi=-\int d^4x\sqrt{-g}\frac{\phi^2}{4G}+{i \over 2} {\rm Tr}
({\cal
D}^{-1}\phi)({\cal D}^{-1} \phi).
\label{effpot}
\end{equation}
Next, we shall calculate the contribution of fluctuations $\phi$ to
the effective action, taking into account the quark loop in the
gravitational field. (Note that this corresponds to the
integral (11) and Fig.1a in \cite{[1]}.) For our purpose of making
estimates of the role of fluctuations, it is sufficient to limit
ourselves to the consideration of fluctuations depending only on
time. In this case, we can extract the necessary kinematical factor
for the meson fluctuation field and then integrate over the
time-component of the loop momentum in the limit of vanishing
external momenta. Finally, after going to the basis for the Dirac
equation in the Einstein universe (see (\ref{19})--(\ref{191}))
we obtain a sum over fermion loop quantum numbers  $l$ instead of an
integration over momenta of free quarks made in \cite{[1]}. The
sum is divergent and we  regularize it by the cut off $\Lambda$. In
this way we obtain the effective action
\begin{equation}
S_\phi=-\int d^4x\sqrt{-g}\left[\frac {\phi^2}{4G}+\frac
{3}{4\pi^2 a^3}
\sum_{l=0}^\infty e^{-\omega_l/\Lambda}
2(l+1)(l+2)\left (-\frac {\phi^2}{E_l}+ \frac
{4\sigma_0^2\phi^2-(\partial_t \phi )^2} {4E_l^3}\right
)\right ],
\label{phiaction}
\end{equation}
where $E_l$ and $\omega_l$ are given in (\ref{190}). The summation
over $l$ in the first term in parenthesis in the above equation
cancels out by the term $\phi^2/4G$, due to the stationarity
condition (\ref{gap})
\begin{equation}
1 ={3G\over \pi^2a^3}\sum_{l=0}^\infty e^{-\omega_l/\Lambda}
2(l+1)(l+2)\frac {1}{E_l}.\label{st}
\end{equation}
After this we obtain 
\[
S_\phi=\int dt {3\over 2}\sum_{l=0}^\infty e^{-\omega_l/\Lambda}
2(l+1)(l+2)\frac {1}{E_l^3}\left ({1\over
    4}\dot\phi^2-\sigma_0^2\phi^2\right),
\]
or $S_\phi=\int dt L_\phi$
with the Lagrange function
\begin{equation}
 L_\phi={1\over 2}\left (\dot\phi^2-4\sigma_0^2\phi^2\right){\cal
 V}{\mathcal Z}^{-1}, 
\label{lagrange}
\end{equation}
where ${\cal V}=\int
d^3x\sqrt{-g}=2\pi^2a^3$ is the space volume, and the renormalization
${\mathcal Z}$-factor is defined as 
\begin{equation}
{\mathcal Z}^{-1}={3\over
   4{\cal V}}\sum_{l=0}^\infty
e^{-\omega_l/\Lambda}
2(l+1)(l+2)\frac {1}{(\omega_l^2+\sigma_0^2)^{3/2}}.
\label{major}
\end{equation}
For comparision, let us consider the limiting case of flat space
which is reached by the replacements
\[
\sum_l \to \int dl,~~~(l+1)(l+2)\to \vec p^2 a^2
\]
in (\ref{major}). Then the ${\mathcal Z}$-factor takes the form of
the integral in the Euclidean spacetime
\[
{\mathcal Z}^{-1}=12\int{d^4p\over (2\pi)^4}{1\over
(p_4^2+\varepsilon_p^2)^2}, 
\]
with $\varepsilon_p$ being the quark energy.
This expression  evidently corresponds to the similar formula for
the ${\mathcal  Z}$-factor in the flat space case considered in
\cite{[1]}. Next, let us perform the field renormalization 
$\phi={\mathcal Z}^{1\over 2} \phi_r$ in (\ref{lagrange}).
Quantum fluctuations of the boson field near the ground state
$\sigma_0$ can now be estimated, if we consider the renormalized
expression (\ref{lagrange}) as the Lagrange function for a harmonic
oscillator (here, we follow the idea of \cite{rubakov}
\footnote{In our case of the NJL model, the consideration of the 
quark loop diagram is essential (see \cite{[1]}). This differs from 
Ref. \cite{rubakov}, where the $\phi^4-$model of a self-interacting
scalar field was considered and the scalar loop contribution to the
fluctuation Lagrangian was calculated.})
with the mass $m$ and frequency $\omega$, formally given here
by the relations
\[ m=
{\cal V},~~~\omega^2=4\sigma_0^2\equiv M^2 .
\]
Then we can estimate the quantum fluctuations as
\[
\langle\phi^2\rangle \approx
{1\over m\omega}= {1\over {\cal V}2\sigma_0}
={1\over {\cal V}M},
\]
where M is the mass of the composite $\sigma$-meson. Thus, we obtain
\begin{equation}
\frac{\sigma^2_0}{\langle\phi^2\rangle}\approx 
a^3\sigma_0^3.
\label{estimate}
\end{equation}
The estimate (\ref{estimate}) gives a criterion for the role of
quantum fluctuations for a system with finite volume.
Clearly, quantum fluctuations can be considered negligible, if
$\sigma_0^2\gg \langle\phi^2\rangle$. This is in agreement with the
physical requirement that quantum fluctuations should be negligible
if $a\sigma_0\gg 1$, i.e., if the radius of the universe is much
greater than the Compton wavelength $\lambda=\frac{1}{M}=
\frac{1}{2\sigma_0}$ of the $\sigma$-meson (quarks) (see also
\cite{rubakov})
\footnote{Note, that there arise also corrections from meson loops to
the quark and pion condensaates and the meson mass $M$, which are
of order $O(1/N_c)$ \cite{fluctuations}. All these corrections 
are surely suppressed in the case of large numbers of colors
$N_c$, where the (induced) coupling constants become small.}.

The above estimates are certainly of a qualitative nature, and hence
they do not allow us to find an exact value of a critical radius
such that symmetry breaking for lower values of the curvature radius
of the Einstein universe is forbidden.

\end{document}